# 10

# Body models in humans, animals, and robots: mechanisms and plasticity

*Matej Hoffmann*

## 10.1 Introduction

Ulric Neisser distinguishes five different selves: the ecological self, the interpersonal self, the extended self, the private self, and the conceptual self (1988). The high-level facets of the self—accessible to consciousness, incorporating linguistic information, etc.—have been receiving relatively more attention. However, here I will focus on the lowest level—the ecological or sensorimotor self or the 'body schema'—which constitutes a key foundation for the rest. The description by Graziano and Botvinick (2002) nicely expresses the sensorimotor, multimodal, and spatial nature of the body representations on which I focus: 'implicit knowledge structure that encodes the body's form, the constraints on how the body's parts can be configured, and the consequences of this configuration on touch, vision, and movement'.

The first key question I want to address is what the fundamental differences are in which animals, humans, and robots represent their bodies. While the main goal is to get an understanding of the mechanisms of 'the body in the biological brain', the 'robot world' can be instrumental here in two ways. First, there is a large body of mature mathematical tools for representing kinematics and dynamics and for employing these representations in movement planning and control, as well as for learning models of physical systems (system identification). These constitute, in some sense, the 'ideal world', a neat mathematical description of the problem, which opens up a useful perspective on the body models that evolution has arrived at. Second, robots can serve as embodied computational models of biological body representations. Humanoid robots possess morphologies—physical characteristics, as well as sensory and motor apparatuses—that are, in some respects, akin to human bodies and can thus be used to expand the domain of computational modelling by anchoring it to the physical environment and a physical body and allowing for instantiation of complete sensorimotor loops.

The second key question is: Which properties of the biological 'body schema' could be transferred to robots to make them more adaptive and resilient? On one hand, robots are endowed with neatly engineered body models and control algorithms. Yet, in many respects, their performance in commanding their bodies in unstructured environments, adapting to failures or tools, etc., is still hugely lagging behind their biological counterparts. Therefore, Section 10.4 will examine which of the characteristics of the 'body in the brain' robots should take on board.

## 10.2 Body models—octopus, primates, robots

Biological and artificial agents have very different bodies, as well as very different representations thereof. In this section, I will look at some of the characteristics of bodies and 'brains' of invertebrates, primates, and robots. Reaching will be used throughout as a behaviour that requires some form of—implicit or explicit—body model.

### 10.2.1 The invertebrate brain and reaching in the octopus

[1]Unlike in vertebrates, invertebrate species show an enormous diversity in body plans and nervous organization (Marder, 2007; Zullo & Hochner, 2011). With more complex bodies and nervous structures, there is a tendency toward centralization with the formation of a structured cephalic ganglion. Ganglia or their groups become larger and tend to form semi-autonomous systems for sensorimotor control. Brain development in the rostral part of the animal comes also from the presence of distal sensing such as vision. Within the higher nervous system, sensory feedback areas tend to be topographically organized; central ganglia receive projections from various body parts and show general somatotopy (Walters et al., 2004; Vitzthum, Muller, & Homberg, 2002; Wong, Wang, & Axel, 2002). Interneurons become more common and constitute a key element in processing and integrating information.

The most advanced invertebrate class is the cephalopods—highly derived molluscs. They feature, on one hand, the highest centralization of the nervous system. On the other hand, next to the central nervous system (CNS) composed of the brain and two optic lobes, there is a large peripheral nervous system (PNS) of the body and the arms. The brain consists of 30–40 interconnected lobes with a high degree of cross-talk; yet, the interconnections appear less elaborate than in vertebrate brains (Young, 1971). Despite the high level of centralization and in contrast to vertebrate and insect brains, there is no obvious somatotopic arrangement in either motor or sensory areas.

The most prominent and most intelligent, and with the largest nervous system among cephalopods, is the octopus. The octopus has a unique embodiment—a flexible body and eight arms with virtually infinite degrees of freedom. Brain stimulation reveals that motor control is hierarchically organized into three functional levels. In the higher motor centres located in the basal lobes, microstimulation evokes complex movements which are, however, not somatotopically represented, but controlled by parallel overlapping circuits representing individual motor programmes. The basal lobes receive inputs from the optic lobes and other sensory centres.

[1] The beginning of this section draws heavily on Zullo & Hochner (2011).

Yekutieli et al. (2005a) and Yekutieli, Sagiv-Zohar, Hochner, & Flash (2005b) developed a dynamic model of the octopus arm and used it to hypothesize the mechanism of how a reaching movement is generated. Despite the complexity of the arm, they found that in their model, only two control parameters suffice to fully specify the extension movement of the arm: (1) the amplitude of the activation signal; and (2) the activation travelling time. This hypothesis seems in line with electromyogram (EMG) recordings and other evidence from the real octopus. Larger amplitudes would result in the same kinematics, but larger forces, increasing the arm's stability against perturbations. For reaching directed at a particular target, two additional control parameters are necessary for the orientation of the arm base. Considering both the experimental and the simulation results, Yekutieli et al. (2005b) speculate that the octopus reaches toward a target using the following strategy:

(1) Initiating a bend in the arm so that the suckers point outward.
(2) Orienting the base of the arm in the direction of the target or just above it.
(3) Propagating the bend along the arm at the desired speed by a wave of muscle activation that equally activates all muscles along the arm.
(4) Terminating the reaching movement when the suckers touch the target by stopping the bend propagation and thus catching the target.

There are three kinematic control parameters (two angles for arm base orientation and one for movement speed) and one dynamic control parameter corresponding to muscle force.

However, the behavioural repertoire of the octopus is greater. Gutnick, Byrne, Hochner, and Kuba (2011) prepared a special setup where the octopus has to guide one of its arms through a maze to reach food in one of three branches marked by a visual cue. The octopus is capable of such 'hand–eye coordination'. Also, instead of the stereotypical largely feed-forward bend propagation, it uses a much slower—but possibly feedback-controlled—'search movement'. Finally, to bring food to the mouth, the arm is bent in a specific way, creating 'virtual joints' along it.

### 10.2.2 The body in the primate brain

It is not in my capacity or my goal to review the structure and function of vertebrate nervous systems. Instead, I will briefly discuss how the body is represented in the brain of primates, which include monkeys, apes, and humans. Again, reaching will serve as an example in which a body model of sorts needs to be employed.

The presence of various 'body maps' in the primate brain has fascinated scientists and the general public alike, spurred by the account of Head and Holmes (1911) and the discovery of the somatotopic representations (the 'homunculi') in the primary motor and somatosensory cortices (Leyton & Sherrington, 1977; Penfield & Boldrey, 1937). Neurological conditions and accounts of a whole range of illusions regarding own body perception (e.g., rubber hand illusion, out-of-body experience, apparition) generated both seminal research articles (e.g., Botvinick & Cohen, 1998; Lenggenhager, Tadi, Metzinger, & Blanke, 2007) and public interest. The attention devoted to the representations of the body in the brain has also led to numerous attempts at describing or defining them, and proposals of a variety of concepts, including superficial and postural schema (Head & Holmes, 1911), body schema, body image, corporeal schema, etc., have been put forth. One characteristic common to all these representations is their multimodal nature—they dynamically integrate information from different sensory modalities (visual, tactile, proprioceptive, vestibular, auditory) (Azañón et al., 2016), while not excluding motor information. However, the concepts of body schema, body image, and many others are umbrella notions for a range of observed phenomena, rather than a result of identification of specific mechanisms. The field is thus in a somewhat 'chaotic state of affairs' (Berlucchi & Aglioti, 2010), with limited convergence to a common view (Graziano & Botvinick, 2002; Holmes & Spence, 2004).

Reaching behaviour in primates bears some similarity to that in the octopus. A reaching movement has some high-level characteristics like the direction of a hand's movement in space, the extent of the movement (amplitude), the overall duration (movement time), and other parameters such as anticipated level of resistance to the movement (Schöner, Tekülve, & Zibner, 2018). Also, movement generation involves cooperation between the CNS and PNS. The exact mechanisms of motor control in humans and other primates are still debated (see, for example, Lisman, 2015). One view—the equilibrium point theory—posits that high-level descending motor commands modulate the peripheral reflex loops (such as the stretch reflex) and set a desired muscle length (Feldman, 2011). Compared to invertebrates, motor control in vertebrates, specifically mammals and in particular primates, becomes more 'cortical' and the motor cortex has the possibility of more direct control over the details of a particular movement, which is likely correlated with the need for dexterous manipulation.

Also, it may not be possible to decouple the movement preparation phase from its execution. The actual movement may be a product of couplings of feed-forward and feedback control that are necessary to understand the effects such as the uncontrolled manifold (Scholz & Schöner, 1999; Martin, Reimann, & Schöner, 2019).

It is also not completely clear whether a prerequisite for a reaching movement is localizing the effector—say the arm/hand—in space first. Through cortical microstimulation, Graziano was able to elicit stereotypical movements in the monkey, irrespective of the starting position (2006). However, Schöner et al. (2018) speculate that the stimulation could drive an update of the hand position first, which would be subsequently used for the movement generation—in line with Scott and Kalaska (1997) who found that neural tuning curves in the motor cortex depend on the arm's kinematic configuration.

In this chapter, I will specifically focus on a task in which localizing the own body cannot be circumvented—reaching to own body parts. For example, Lisman (2015) notes: 'More difficult to imagine is how one can reach one's ear lobe without visual guidance. This requires knowledge of the position of the target and the position of the limbs. It has only recently become clear that there is indeed a population code that represents hand position (Hauschild et al., 2012), but surprisingly little is known about the muscle and joint signals that allow this computation to be made (Weber et al., 2011).'

### 10.2.3 Robot body models

The world of robots and their body models is completely different. The first striking difference is that describing it is significantly easier; robot body representations and control schemes are designed by engineers and are thus very transparent and, unlike in the brain, we have complete access to all information. Let us take the iCub humanoid robot as an example. The iCub (Metta et al., 2010) (see Figure 10.1) is a baby humanoid robot that was designed after a 4-year-old child—with similar body proportions, kinematic structure, and sensorimotor capacities. At the same time, it is, to a large extent, a product of (great) engineering.

In Figure 10.1A, there is a cartoon of the robot, side by side with its computer-aided design (CAD) model (see Figure 10.1B) and the basic kinematic structure (joints and links) (see Figure 10.1C). Complete knowledge of the robot structure can be used to obtain a mathematical description of the robot's kinematics. This essentially corresponds to a sequence of coordinate transformations between all the reference frames in Figure 10.1D. Every such transformation consists of three translations and three rotations and

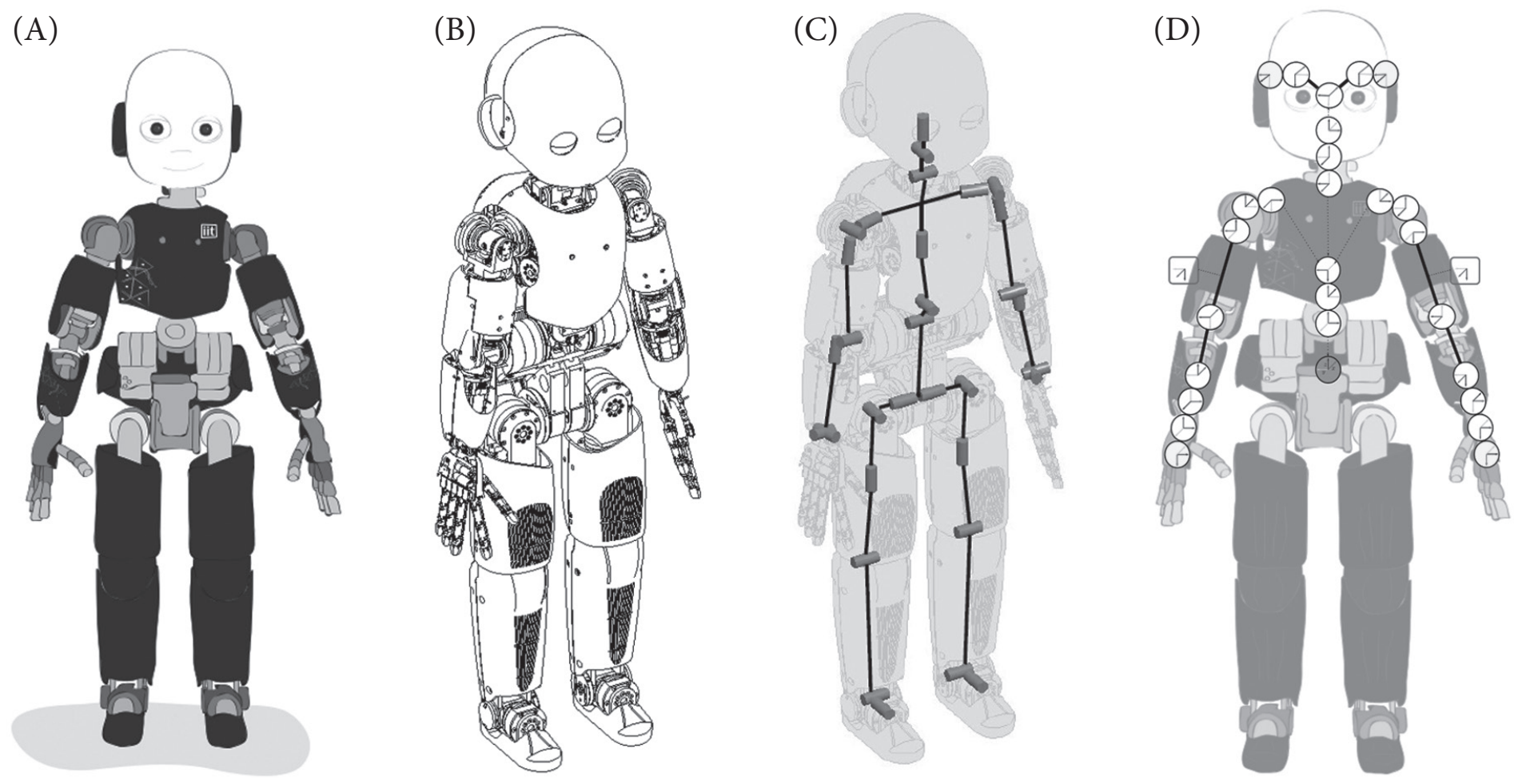

**Figure 10.1** The iCub humanoid robot. (A) Cartoon of the robot. (B) CAD model. (C) Kinematic structure. (D) Reference frames in upper body.

*Image credits*: (A) iCub cartoon: Reproduced courtesy of Laura Taverna, Italian Institute of Technology. (B) (C) iCub kinematic structure: Reproduced with permission from (Parmiggiani, et al., 2012), and courtesy of Alberto Parmiggiani. (D) Reproduced courtesy of Jorhabib Eljaik.

has thus six degrees of freedom (DoFs). However, as the robot structure is subject to specific constraints in three-dimensional space, four parameters suffice to characterize the transformation between consecutive links/joints. This is the essence of the Denavit-Hartenberg convention in which every link $i$ (imagine getting from the elbow to the wrist along the forearm, for example) is described by four parameters: two lengths $a_i$ and $d_i$, and two angles $\alpha_i$ and $o_i$. In the iCub—and in most robots for that matter—all joints are revolute and have a single rotation axis, i.e., a single rotational DoF, like the human elbow (a 'hinge joint'). Figure 10.2 shows a schematic of this representation for the upper body of the robot.

### 10.2.3.1 Forward kinematics and inverse kinematics

The robot model describes the fixed characteristics of the robot body (long-term or 'offline' body representation; see later). The model in Figure 10.2 can be transformed into equations, whereby the coordinate transformation needed to get from one link to the next (e.g., from elbow to wrist) can be obtained as a simple matrix multiplication. To go over more links/joints (e.g., from torso to hand), these multiplications are simply sequenced. In the canonical form, this will only work for a single posture of the robot, like the one in Figure 10.1. To know where the body currently is in space, it has to be combined with the robot 'proprioception'—the joint angle values. These are plugged into the equations as one of the orientation parameters ($o_i$). This operation—combining current joint angle values with the robot model—is known as *forward kinematics*, which provides a mapping from *joint space* to *Cartesian space* (also called

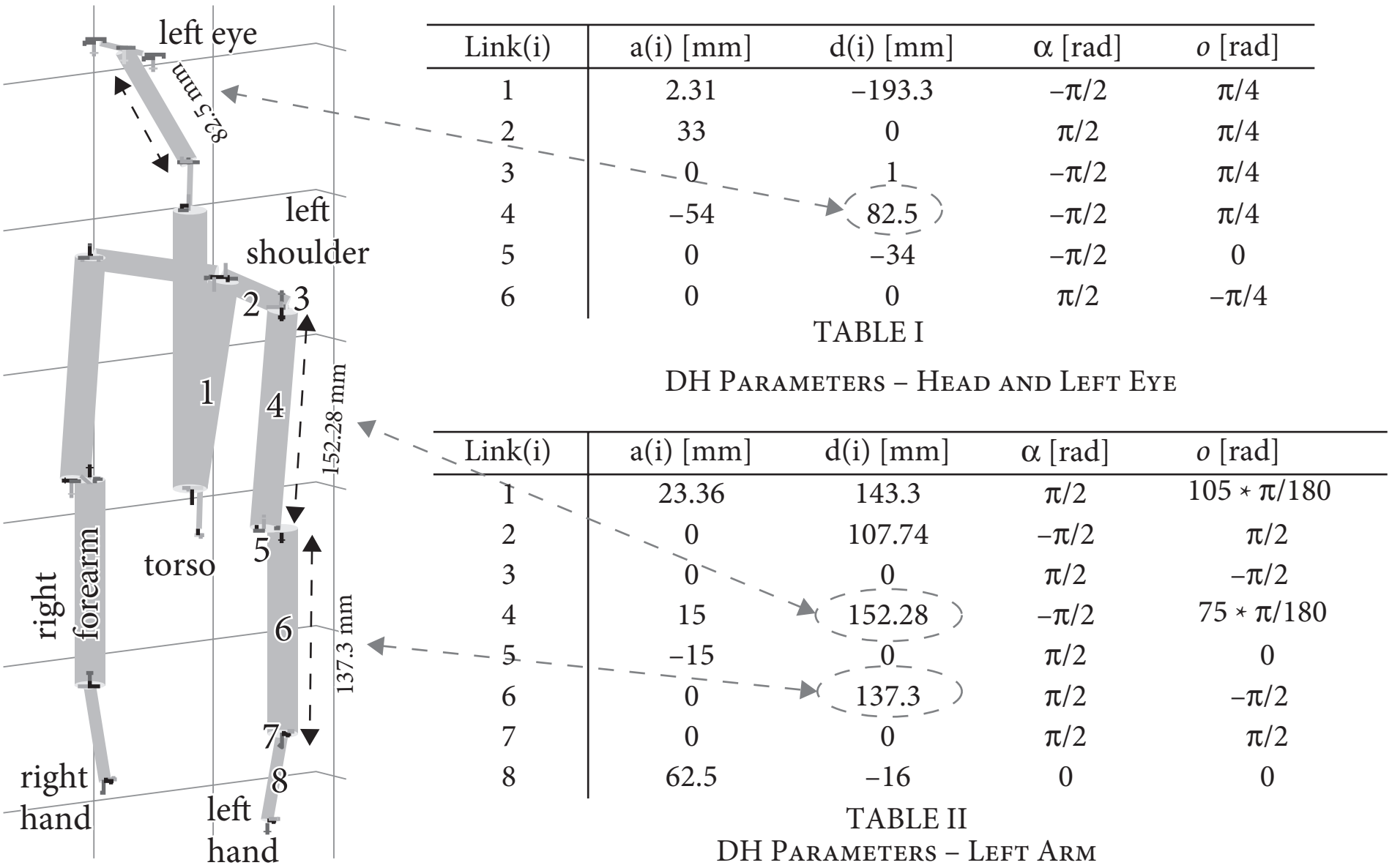

| Link(i) | a(i) [mm] | d(i) [mm] | α [rad] | o [rad] |
|---|---|---|---|---|
| 1 | 2.31 | −193.3 | −π/2 | π/4 |
| 2 | 33 | 0 | π/2 | π/4 |
| 3 | 0 | 1 | −π/2 | π/4 |
| 4 | −54 | 82.5 | −π/2 | π/4 |
| 5 | 0 | −34 | −π/2 | 0 |
| 6 | 0 | 0 | π/2 | −π/4 |

TABLE I

DH PARAMETERS – HEAD AND LEFT EYE

| Link(i) | a(i) [mm] | d(i) [mm] | α [rad] | o [rad] |
|---|---|---|---|---|
| 1 | 23.36 | 143.3 | π/2 | 105 * π/180 |
| 2 | 0 | 107.74 | −π/2 | π/2 |
| 3 | 0 | 0 | π/2 | −π/2 |
| 4 | 15 | 152.28 | −π/2 | 75 * π/180 |
| 5 | −15 | 0 | π/2 | 0 |
| 6 | 0 | 137.3 | π/2 | −π/2 |
| 7 | 0 | 0 | π/2 | π/2 |
| 8 | 62.5 | −16 | 0 | 0 |

TABLE II

DH PARAMETERS – LEFT ARM

**Figure 10.2** iCub kinematic model—upper body. (Left) Matlab visualization of the kinematic chains and reference frames. (Right) Denavit-Hartenberg (DH) parameters for the head and left eye and left arm. Correspondences for certain link lengths are marked.

*operational space* or *task space*). Thus, transformations between, say, hand frame, body frame, or eye frame can be readily obtained.

For a robot to reach to a specific position and with a specific orientation in Cartesian space—no matter in which reference frame the target is expressed as it can be transformed to the base frame—an inverse mapping is needed, from Cartesian space to joint space, i.e., to acquire the joint angles of the robot arm when the end effector (the hand) contacts the target. This is dealt with by *inverse kinematics*. Unlike forward kinematics, this is a harder problem and not just a matter of substituting for current joint angle values into the equations. Reaching for a target in three dimensions with a specific orientation constraints the robot pose in six dimensions (three positions and three orientations). Hence, a minimum of six DoFs—six joints—is required on the robot part. For six DoF manipulators—robot arms with six rotational joints—that have a specific geometric structure, a closed-form solution can be obtained. That is, a solution can be found instantaneously. However, in general, one has to resort to numerical methods. Robots with more than six DoFs have multiple ways of reaching for the target and hence are called *redundant manipulators*. Additional criteria are needed to choose among the solutions. In the iCub, there are seven DoFs in every arm and three additional ones in the torso. The manipulator is thus highly redundant. Inverse kinematics is solved numerically by using a non-linear optimizer (Pattacini, Nori, Natale, Metta, & Sandini, 2010).

#### 10.2.3.2 Motion control

Inverse kinematics provides the joint space configuration for the robot in the final position at the target. However, it does not automatically deal with the trajectory—joint and end effector positions in time—needed to move between the initial and final positions. *Trajectory generation* constitutes its own discipline that deals with planning such smooth movements using different interpolation methods, for example. Once the trajectory in joint space is planned, low-leveI motor controllers in every joint can be used to bring about the desired movements in time. In the iCub, Pattacini et al. (2010) designed a bio-inspired dynamical systems-based controller that produces smooth, minimum jerk trajectories in which the end effector (the hand) follows a quasi-straight line.

#### 10.2.3.3 Dynamics

While kinematics deals with positions, velocities, and accelerations, dynamics deals with equations of motion and forces that are needed to produce a particular movement. The reader is referred to any robotics textbook on the topic. For the iCub, Nori et al. (2015) provide an example.

## 10.3 Characteristics of body representations

In this section, I will attempt to compare the most important features of body representations in animals, humans, and robots. First, a note on terminology is in order. I use 'body representations' as a general concept that should encompass both body schema and body image and possibly other body-related notions. However, by using the word 'representation', it is not my goal to take a stance in the philosophical debate on representationalist versus sensorimotor approaches to body awareness (de Vignemont, 2015). The account will admittedly be biased toward the 'representationalist' viewpoint (e.g., Carruthers, 2008; Longo et al., 2010)—also because I come from robotics and computer science—and will not address the phenomenological perspective and the 'lived body' (Merleau-Ponty, 1945). At the same time, I also fully endorse the 'sensorimotor approach' and I want to avoid, or at least reflect upon, imposing the representationalist stance typical of robotics and (good old-fashioned) artificial intelligence (Haugeland, 1985) onto the biological 'body in the brain' (see also our attempt in Hoffmann et al. (2017)). Webb (2006) provides a useful clarification of the terms *transformation*, *encoding*, and *representation*. The word r*epresentation* should be reserved to the strong notion of standing in for something—properties or states of the body that can be manipulated also when the body itself cannot be used or sensed directly. Sometimes, the body can be used directly—imagine the reaching in the octopus discussed above—and it is probably more natural to think that the 'brain is in the body' and does not have to have it all represented, rather than the 'body in the brain' is embodied (cf. a discussion in Alsmith & de Vignemont (2012)).

According to de Vignemont (2015), there are three principal *taxonomies of body representations*. I will list them below, together with their relationship to the account in this book section:

(1) The triadic taxonomy based on the 'format' of body representations (Schwoebel & Coslett, 2005) distinguishes:
   (a) Sensorimotor body representation (also known as body schema).
   (b) Visuo-spatial body representation (or body structural description).
   (c) Conceptual body representation (or body semantics).

   My account will span roughly the first two, leaving the conceptual body representation aside.
(2) The functional dyadic taxonomy (Dijkerman & De Haan, 2007; Gallagher, 2006; Paillard, 1999), based on the perception–action model of vision ('ventral stream' or 'what' versus 'dorsal stream' or 'where/how') (Milner & Goodale, 2006), distinguishes:
   (a) Body schema—sensorimotor representations of the body used for action planning and control.
   (b) Body image—lacking a unifying positive definition (Berlucchi & Aglioti, 2010; de Vignemont, 2010); comprises all the 'other' (than body schema) representations about the body that are not used for action: perceptual, conceptual, or emotional.

   My account will focus on the representations for action. Robot body models are also primarily geared toward action. However, interestingly, as their models are engineered from the outside, they do carry a lot of the flavour that is typical rather of the 'what' pathways that care about semantics of objects, etc.

(3) The temporal dyadic taxonomy (Carruthers, 2008) is based on the dynamics of body representations and contrasts:
  (a) Long-term or 'offline' body representations, such as limb size—what the body is usually like. These are relatively stable in adulthood and may include some innate components about body structure (e.g., two arms and two legs). Carruthers (2008) also argues 'that the offline body representation must be an integrated representation, a failure to integrate leads to body integrity identity disorder.' 'That is, it must represent the body as a single thing, rather than a collection of parts.'
  (b) Short-term or 'online' representations of the body as it is currently, such as in which posture, constantly updated on the basis of afferent and efferent information.

  In this case, both are obviously equally relevant. From the robotics perspective, the *long-term body representation* can be equated with *body model*. The short-term representation would be the *body state*. I will focus more on the long-term body representations.

To give the discussion concrete contours, let us look at the example in Figure 10.3. Figure 10.3A depicts a scenario from a series of studies on infants where they were observed reaching for a vibrotactile stimulus (buzzer) (Hoffmann et al., 2017; Leed, Chinn, & Lockman, 2019). Two components seem necessary: (1) localizing the stimulus on the body; and (2) reaching for it. The centre in Figure 10.3A depicts a possible decomposition into blocks. The left part (up to the 'localization' blocks) (Tamè, Azañón, & Longo, 2019) deals with the 'localization' or somatoperceptual processing part. Once the target is localized (*spatial localization of touch* block), it can be adopted as the *reaching target* (illustrated with a '~'), and the motor action can be prepared and executed. The different shapes for the blocks were reproduced after (Longo et al., 2010) as follows: 'inputs are depicted as diamond shapes, body representations as ovals, and perceptual processes as rectangles'. It is interesting to look at this from the temporal taxonomy perspective; it seems that only the long-term or offline modules count as representations here. The percepts, however, represent 'states of the body' and can be viewed as short-term or online body representations. Heed, Buchholz, Engel, and Röder (2015) offer a different conceptualization of this scenario, also known as *tactile remapping*—the transformation of a coordinate in a skin-based reference frame into a coordinate in an external reference frame by integration of posture information. Heed et al. actually present three variants of the schematics (not reproduced here): (a) remapping view; (b) integration view; and (c) sensorimotor contingency view. In (b) and (c), the arrows between the 'localization' part and the 'action' part point in both directions, which illustrates another important aspect of the problem—a sequential processing view and decoupling of the perception and action parts may not be justified, something we touched upon in Section 10.2.2. Analysis of the infant reaching behaviour in the buzzer experiments shows that looking at the target and reaching often happen simultaneously; in addition, for targets on the arms/hands, both the limb with the target and

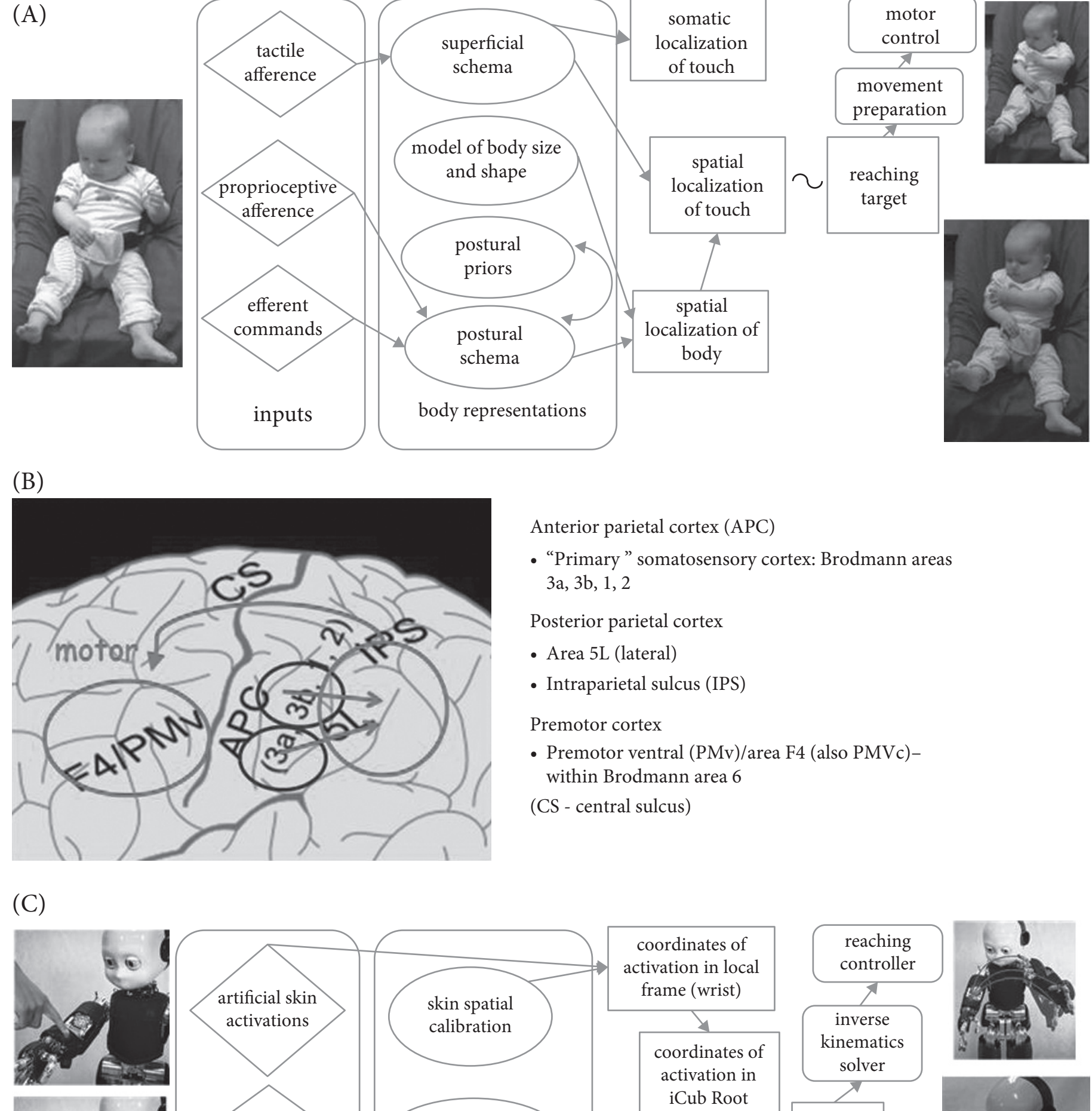

**Figure 10.3** Reaching to the body. (A) Scenario with reaching for vibrotactile stimuli in infants (Hoffmann et al., 2017). Photos show an infant presented with the stimulus (left), in the process of reaching (top right), and in the final posture (bottom right). Centre: a conceptual model of somatosensory processing (left part, up to '~', adapted from Longo, Azañón, and Haggard (2010) and Tamè, Azañón, & Longo, (2019)) and reaching action. (B) Schematic illustration of cortical areas that may be responsible for bringing about the behaviour. (C) Similar scenario on the iCub humanoid robot: tactile stimulus (left), motor action (top right), and final configuration (bottom right). Centre: block diagram illustrating the modules used to generate the 'self-touch' behaviour (Roncone, Hoffmann, Pattacini, & Metta, 2014).

*Image credits*: (A) Flowchart adapted from (Tamè, Azañón, & Longo, 2019) under a Creative Commons Attribution License (CC BY) (https://creativecommons.org/licenses/by/4.0/) and with permission from Longo, M. R., Azañón, E., and Haggard, P. (2010). With additional blocks illustrating the motor part. (B) Brain image adapted from Hugh Guiney/Creative Commons Attribution-Share Alike 3.0 Unported license (CC BY- SA 3.0) (labels added). (C) Reproduced courtesy of Alessandro Roncone and Matej Hoffmann.

the reaching contralateral arm often move simultaneously (Chinn, Hoffmann, Leed, & Lockman, 2019). The more general point of recurrent connections also relates to the state estimation problem; the 'percepts' may not be the result of a single pass that combines information from different sources but may be the result of conflict reconciliation where activations need to flow back and forth.

Figure 10.3B schematically illustrates some of the cortical areas of the monkey brain that may be involved in generating this behaviour. Finally, Figure 10.3C shows an instantiation of a similar scenario on the iCub humanoid robot. There is one important difference—this is not a conceptualization of the behaviour; this is an actual pipeline that has been used in Roncone, Hoffmann, Pattacini, and Metta (2014) (for the video, see https://youtu.be/pfse424t5mQ). The blocks correspond to actual pieces of software. Surprisingly, there is quite a good match with the schematics in Figure 10.3A, which may be because the schematics of Longo and colleagues is somewhat classical and thus compatible with engineering models that—mostly for practical reasons—typically follow a modular design and 'sense–think–act' logic.

There are almost countless characteristics of body representations that we can think of. In what follows, I will sketch some of the important ones, focusing in particular on those where contrasting the biological and robot worlds can bring the most insight. Hence, I will, for example, leave the unconscious versus the conscious aspect aside—as discussed above, the focus here is on the sensorimotor level. I will take a number of examples from biology and robotics, developing the ideas in Hoffmann et al. (2010) and Hoffmann & Pfeifer (2018). I will also, sometimes quite speculatively, attempt to chart the body schema and body image onto the hypothetical axes. Body image will largely stand for visuo-spatial representation of the body or body structural description—body percept rather than body concept. The brain areas involved are also only schematically illustrated.

In Figure 10.4, the iCub humanoid robot and the kinematic model of its upper body is depicted in panel (A). The model has been essentially handcrafted by following the Denavit-Hartenberg convention and supplying the corresponding lengths and angles from the CAD model of the robot. Panel (B) adds the possibility to calibrate the same model automatically as the robot exploits self-observation and self-touch configurations (Stepanova, Pajdla, & Hoffmann, 2019). Panels (D) and (E) depict other examples of robot self-calibration. Bongard, Zykov, and Lipson (2006) used a quadrupedal machine continuously 'self-modelling' itself. The robot model had a special nature; it consisted of a physics-based simulator with a copy of the robot's limbs, motors, sensors, and even the environment. Within this engineered 'world and body model', the robot would search for its kinematic structure by comparing the actions and their sensory consequences from the physical world with those in the simulator. Sturm, Plagemann, and Burgard (2009) had a robot arm observe 'itself' using a camera and infer its model—learning the structure of a Bayesian network—from motor actions and observations in the camera. Panel (C) schematically illustrates humans and the body schema and body image.

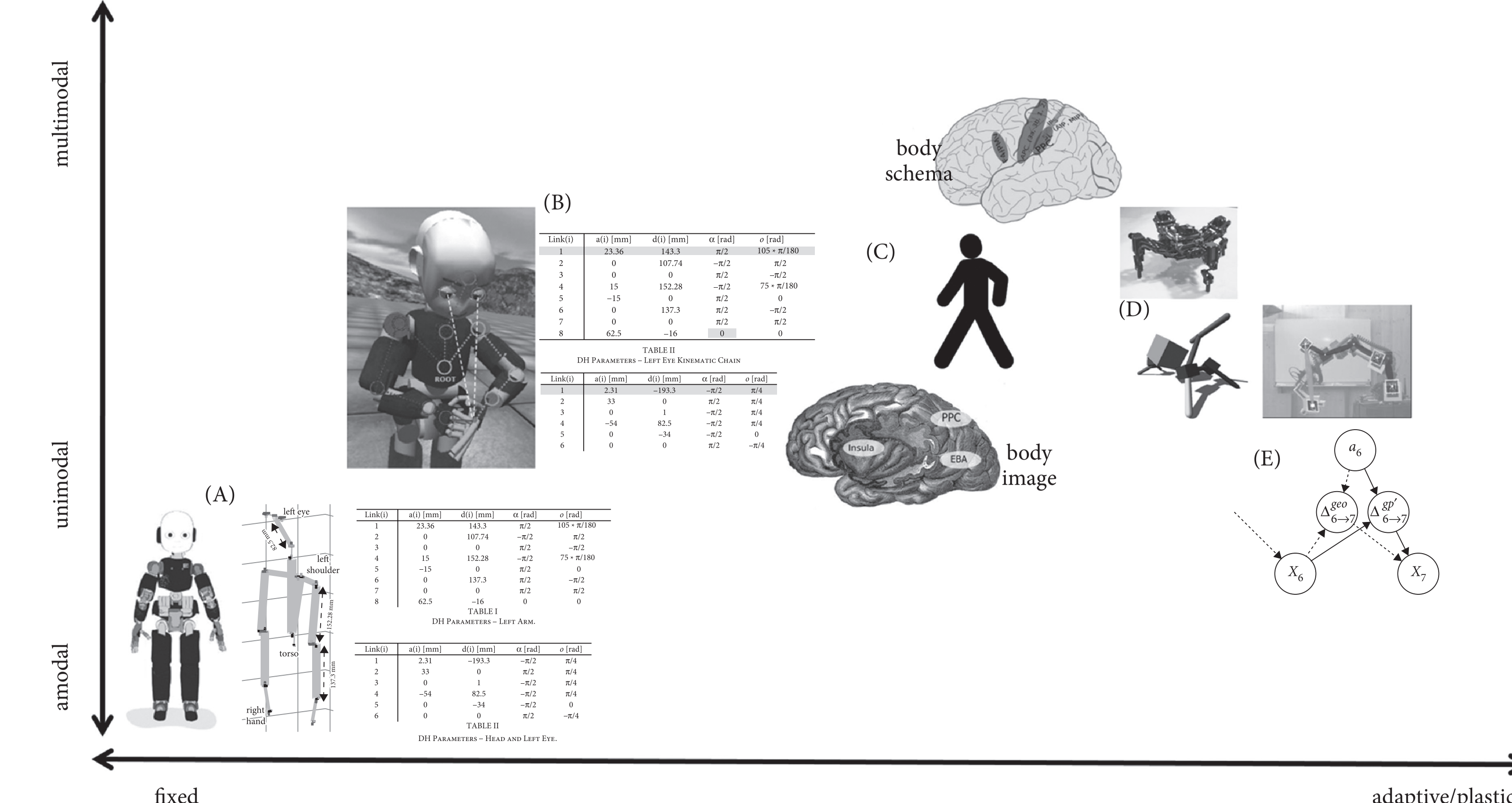

| Link(i) | a(i) [mm] | d(i) [mm] | α [rad] | o [rad] |
|---|---|---|---|---|
| 1 | 23.36 | 143.3 | π/2 | 105 * π/180 |
| 2 | 0 | 107.74 | −π/2 | π/2 |
| 3 | 0 | 0 | π/2 | −π/2 |
| 4 | 15 | 152.28 | −π/2 | 75 * π/180 |
| 5 | −15 | 0 | π/2 | 0 |
| 6 | 0 | 137.3 | π/2 | −π/2 |
| 7 | 0 | 0 | π/2 | π/2 |
| 8 | 62.5 | −16 | 0 | 0 |

TABLE I
DH Parameters – Left Arm.

| Link(i) | a(i) [mm] | d(i) [mm] | α [rad] | o [rad] |
|---|---|---|---|---|
| 1 | 2.31 | −193.3 | −π/2 | π/4 |
| 2 | 33 | 0 | π/2 | π/4 |
| 3 | 0 | 1 | −π/2 | π/4 |
| 4 | −54 | 82.5 | −π/2 | π/4 |
| 5 | 0 | −34 | −π/2 | 0 |
| 6 | 0 | 0 | π/2 | −π/4 |

TABLE II
DH Parameters – Head and Left Eye.

| Link(i) | a(i) [mm] | d(i) [mm] | α [rad] | o [rad] |
|---|---|---|---|---|
| 1 | 23.36 | 143.3 | π/2 | 105 * π/180 |
| 2 | 0 | 107.74 | −π/2 | π/2 |
| 3 | 0 | 0 | π/2 | −π/2 |
| 4 | 15 | 152.28 | −π/2 | 75 * π/180 |
| 5 | −15 | 0 | π/2 | 0 |
| 6 | 0 | 137.3 | π/2 | −π/2 |
| 7 | 0 | 0 | π/2 | π/2 |
| 8 | 62.5 | −16 | 0 | 0 |

TABLE II
DH Parameters – Left Eye Kinematic Chain

| Link(i) | a(i) [mm] | d(i) [mm] | α [rad] | o [rad] |
|---|---|---|---|---|
| 1 | 2.31 | −193.3 | −π/2 | π/4 |
| 2 | 33 | 0 | π/2 | π/4 |
| 3 | 0 | 1 | −π/2 | π/4 |
| 4 | −54 | 82.5 | −π/2 | π/4 |
| 5 | 0 | −34 | −π/2 | 0 |
| 6 | 0 | 0 | π/2 | −π/4 |

**Figure 10.4** Body representation characteristics I: plasticity and number of modalities. (A) iCub humanoid robot and its kinematic model. (B) iCub kinematic self-calibration using self-observation and self-touch (Stepanova, Pajdla, & Hoffmann, 2019). (C) Human and schematic illustration of brain areas important for body representations: body schema (see Figure 10.3B for details) and body image (posterior parietal cortex, PPC; extrastriate body area, EBA; insula) (after Berlucchi & Aglioti, 2010). (D) Four-legged machine learning its body structure (Bongard, Zykov, & Lipson, 2006). (E) Robot manipulator learning body structure from self-observation (Sturm, Plagemann, & Burgard, 2009).

*Image credits*: Figure revised and expanded from (Hoffmann & Pfeifer, 2018). (A) iCub cartoon: Reproduced courtesy of Laura Taverna, Italian Institute of Technology. (B) Reproduced with permission from (Stepanova, Pajdla, & Hoffmann, 2019) (C) Walking human: Public domain (https://commons.wikimedia.org/wiki/File:BSicon_WALK.svg). Body schema brain adapted from Hugh Guiney/ Attribution- ShareAlike 3.0 Unported (CC BY- SA 3.0). Body image brain: Public domain (https://commons.wikimedia.org/wiki/File:Gray731.png). (D) Reproduced with permission from (Bongard, Zykov, & Lipson, 2006), and courtesy of Josh Bongard, University of Vermont. (E) Reproduced with permission from (Sturm, Plagemann, & Burgard, 2009), and courtesy of Jürgen Sturm.

### 10.3.1 Plasticity

Let us first consider plasticity or adaptivity of body models. Traditional robot models are fixed—an industrial manipulator is shipped with its model. It may not be directly available to the customer, but it will be embedded in the robot controller which needs the robot model(s) for operation. Even industrial robots may require occasional (re)calibration, which can be performed using different routines. Less accurate or more complex robots, such as humanoids, may need recalibration more frequently. The work of Stepanova et al. (2019) (see Figure 10.4B) is one of many examples in which the robot kinematic model parameters are calibrated. The approach is rather straightforward; redundant information about the positions of certain body parts—from self-touch or self-observation in this case—drives learning; with two hands physically touching and both cameras observing the point, any mismatch between the position of the corresponding point—after remapping into a common frame of reference—generates an error term used to update the model (all but the grey parameters are calibrated). Body representations in primate brains (Figure 10.4C) are known for their plasticity on several timescales. First, body models need to be discovered by the brain, starting already in the fetal period (see the work of Kuniyoshi and colleagues on embodied computational models of this in Section 10.4). Second, body models need to adapt, as the body grows, for example. Third, they adapt or optimize when some body parts are frequently used in a specific task—when playing a musical instrument, for example. Fourth, they adapt also on very short timescales like when adapting to tools or when the subject is tricked by some of the numerous illusions, like the Pinocchio illusion. In the last case, it is rather a case of 'short-term body schema' adaptation—a state estimation process—rather than adaptation of the body model itself, although this may be happening simultaneously and some after-effects observed. Next to the temporal taxonomy (online versus offline body representations), I will speculate about the distinction between body schema and body image on the plasticity axis. It seems that most of the rapid recalibrations pertain to the body schema which draws more directly on the inputs from different modalities and their integration. Taking the rubber hand illusion (Botvinick & Cohen, 1998) as an example, the body schema adaptation is manifested in the proprioceptive drift; however, the participants also start to quickly incorporate the rubber hand into their bodies—'own' the rubber hand—which is typically associated with body image (and the insula). The suggested primacy of body schema over body image (Pitron, Alsmith, & de Vignemont, 2018) may also be relevant here—body schema adaptation may, in a second step, propagate to body image.

It seems that the brain is rather 'liberal' about the constraints imposed on the models and can be led into believing highly improbable things, like the 'nose elongation' during Pinocchio illusion. Robot models, on the other hand, tend to have quite strict constraints or bounds on the model parameters—capitalizing on the knowledge available from the outside—and would thus not fall for the illusions easily. At the same time, there are limits to the plasticity. Important evidence suggestive of innate and fixed—immune to experience—components of body models comes from the phantom limb phenomenon, which may be experienced following amputation, but also even in some subjects who congenitally lack limbs (e.g., Ramachandran & Blakeslee, 1998). That is, the basic body layout may be, to some extent, hard-wired in the model and immutable. The self-calibrating robots in panels (D) and (E) in Figure 10.4 move in that sense beyond this, as they are able to learn any topology of their body layout—that is why they are positioned more to the right on the plasticity axis. However, this is clearly an oversimplification—these robots surpass the brain plasticity in this single aspect only.

### 10.3.2 Multimodal nature of body models

Standard robot models are *amodal*—they do not depend on any sensory modality; they directly describe physical reality like the geometry of the body (see Figure 10.2 or Figure 10.4A). This holds in some sense also for all the other robot body models in Figure 10.4B, D, E. In Figure 10.4B, the robot model itself is identical to that in Figure 10.4A. The sensory modalities—proprioception, touch, and vision in this case—are needed to collect the redundant information about the body's position in space and update the model. This layer is separated from the model of the robot geometry itself. In Figure 10.4D, E, the situation is quite similar. Similarly to Figure 10.4B, Figure 10.4E features self-observation. In Figure 10.4D (Bongard et al., 2006), three modalities—touch, tilt, and clearance—are used to compare their values from the real robot with those from alternative body layouts in the simulator. Based on this, the case studies are localized on the 'modality' axis. Body representations in the brain are famous for their 'multimodality'. Azañón et al. (2016) review the multisensory contribution to body representations: visual, somatosensory (tactile and proprioceptive), vestibular, auditory, and nociceptive. Hence, 'body in the brain' scores highest on the 'multimodality axis' in Figure 10.4. A divide-and-conquer approach is used throughout the article by Azañón et al.—modality by modality. This is also discussed by the authors—such an approach is useful experimentally, but implausible in reality as there would be a lot of 'cross-talk' between the modalities. Indeed, the body representations are assumed to be, in some sense, unified or coherent. In light of the works on robot self-calibration, this begs the question of whether the brain arrives, in some sense, to an amodal, modality-independent model of the, say, body in space, onto which different sensory modalities converge. This resembles the *emulation theory of representation* proposed by Grush (2004) who uses the Kalman filter[2] metaphor—the amodal, long-term body model is a central representation, which also includes its relationship with individual sensory modalities. The filter can then perform state estimation—current state of the body—by executing a 'sensory update'.[3] To what extent this would be the case for the brain

[2] Coming from signal processing, a Kalman filter is an efficient recursive filter—device or process that removes some unwanted components or features from a signal—that estimates the internal state of a linear dynamic system from a series of noisy measurements.

[3] Grush, taking human arm as an example, contrasts an internal model in the form of a look-up table storing previous input-output sequences with an *articulated model*—a model that includes some variables corresponding to their counterparts in the musculoskeletal system (e.g., elbow angle, arm angular inertia, tension on quadriceps). Some of these variables can be measured (e.g., by stretch receptors) and these sensors can also be simulated in the emulator.

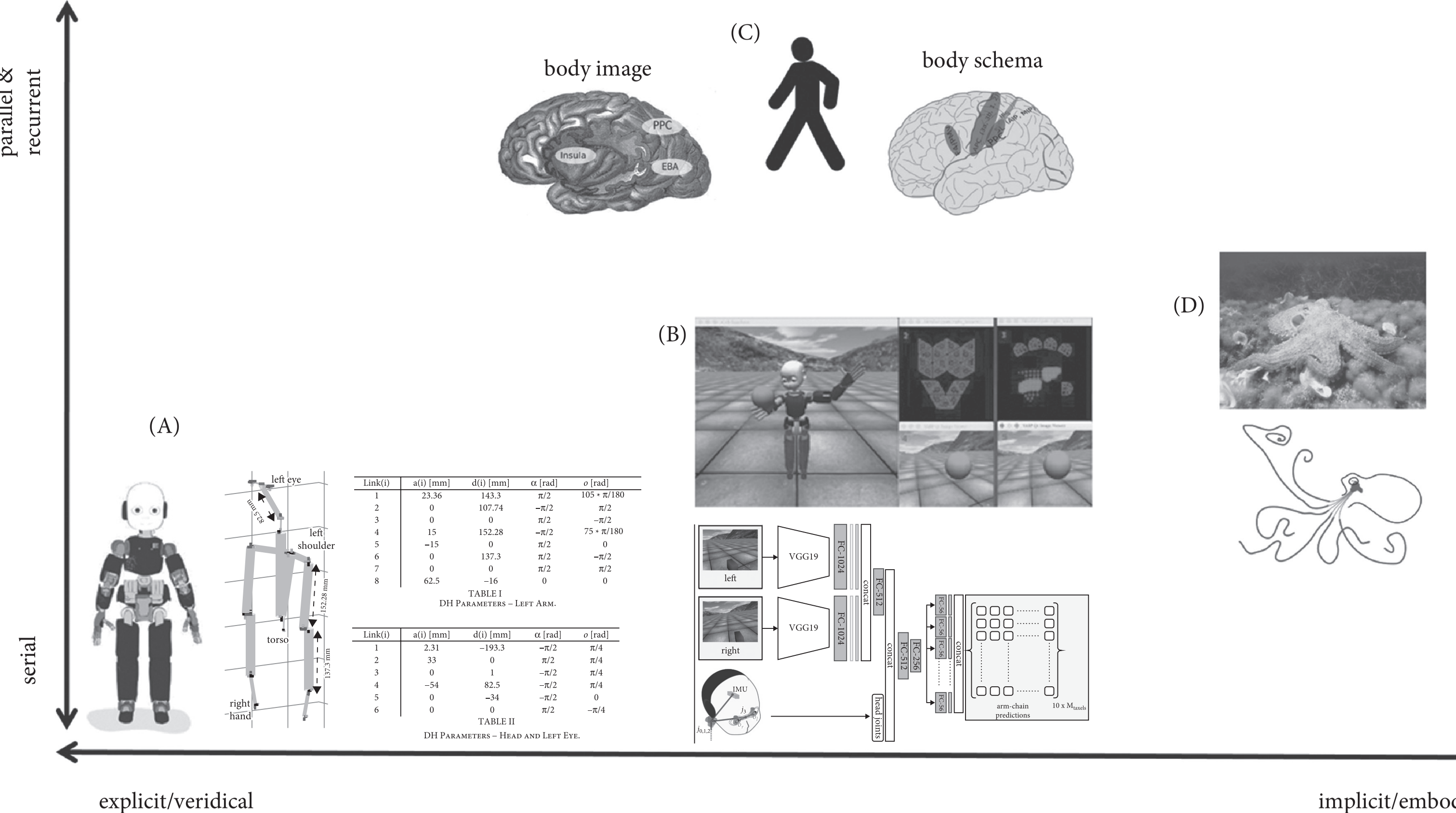

| Link(i) | a(i) [mm] | d(i) [mm] | α [rad] | o [rad] |
|---|---|---|---|---|
| 1 | 23.36 | 143.3 | π/2 | 105 * π/180 |
| 2 | 0 | 107.74 | −π/2 | π/2 |
| 3 | 0 | 0 | π/2 | −π/2 |
| 4 | 15 | 152.28 | −π/2 | 75 * π/180 |
| 5 | −15 | 0 | π/2 | 0 |
| 6 | 0 | 137.3 | π/2 | −π/2 |
| 7 | 0 | 0 | π/2 | π/2 |
| 8 | 62.5 | −16 | 0 | 0 |

TABLE I
DH Parameters – Left Arm.

| Link(i) | a(i) [mm] | d(i) [mm] | α [rad] | o [rad] |
|---|---|---|---|---|
| 1 | 2.31 | −193.3 | −π/2 | π/4 |
| 2 | 33 | 0 | π/2 | π/4 |
| 3 | 0 | 1 | −π/2 | π/4 |
| 4 | −54 | 82.5 | −π/2 | π/4 |
| 5 | 0 | −34 | −π/2 | 0 |
| 6 | 0 | 0 | π/2 | −π/4 |

TABLE II
DH Parameters – Head and Left Eye.

**Figure 10.5** Body representation characteristics II: explicit versus implicit and serial versus parallel. (A) iCub humanoid robot and its kinematic model. (B) iCub learning to reach using deep learning. (C) Human and schematic illustration of brain areas important for body representations: body schema (see Figure 10.3B for details) and body image (posterior parietal cortex, PPC; extrastriate body area, EBA; insula) (after Berlucchi & Aglioti, 2010). (D) The octopus and schematic of its nervous system.

*Image credits*: (A) iCub cartoon: Reproduced courtesy of Laura Taverna, Italian Institute of Technology. (B) Reproduced with permission from (Nguyen, Hoffmann, Pattacini, & Metta, 2019). (C) Walking human: Public domain (https://commons.wikimedia.org/wiki/File:BSicon_WALK.svg). Body schema brain adapted from Hugh Guiney/ Attribution- ShareAlike 3.0 Unported (CC BY- SA 3.0). Body image brain: Public domain (https://commons.wikimedia.org/wiki/File:Gray731.png). (D) Photo: Common octopus- albert kok/CC BY-SA (https://creativecommons.org/licenses/by-sa/3.0). Octopus nervous system: Jean-Pierre Bellier/CC BY-SA (https://creativecommons.org/licenses/by-sa/4.0).

remains unclear. Different distortions of body perception—inheriting some properties of the imperfect representations of individual modalities like the somatosensory homunculus—seem to suggest that the brain has not synthesized a perfect amodal model of its body (see, for example, Fuentes, Longo, & Haggard, 2013 and Longo, 2015). Finally, we can probably say that the body schema is more strongly multimodal, while the body image—at least as a pictorial description of the body based on mainly visual exteroception—would be less multimodal.

### 10.3.3 Explicit/veridical versus implicit/embodied body models

Robot body models are explicit; it is clear what in the model corresponds to what in the body (e.g., a certain parameter to the length of the left forearm) (see Figure 10.2 and Figure 10.5A). They are also objective and veridical; the parameters should be the true physical values of the quantities (lengths, angles, masses, etc.). In the biological realm, representations in general are not like that. Of course, this depends on the school of cognitive science, but there seems to be growing consensus about the embodied and action-oriented nature of cognition (e.g., Engel, Maye, Kurthen, & König, 2013). This should hold for representations of the body as well (Alsmith & de Vignemont, 2012). 'What the nervous system needs to do, in general, is to transform the input into the right action' (Webb, 2006). We can take again the example of reaching behavior. As discussed in Section 10.2.1, the octopus is able to reach for visual targets, but it may not know—and may not need to know—how long its arm is or where it is exactly in space. Orienting the base of the arm and propagating the bend until contact is detected by the suckers may well suffice. This is *embodied action* and may be in line with the fact that no specific body representation—somatotopic or other—sites were found in its nervous system. The need to represent the body, its state, and the complex inverse kinematics and dynamics has been largely offloaded to embodiment—the properties of the muscular hydrostat, supported by the PNS and low-dimensional inputs from the CNS. This is also in line with the thinking of Cisek and Kalaska (2003) who highlight the importance of the online, dynamically generated character of movement generation. At the same time, they also point out that due to conduction delays inherent to the sensorimotor system, purely feedback control is limited, or at least slow, perhaps manifested in the octopus experiments of Gutnick et al. (2011). Successful action is also the only criterion for the 'quality' of what is represented about the animal's body in its brain; there is no need for any objective or veridical representation. Hence, the octopus has been positioned on the far right of the x-axis in Figure 10.5. Insects would be even farther right on this axis, with their 'body in the nervous system' so implicit that we may be reluctant to call this representation altogether.

Similar arguments hold for primate brains, but to a lesser extent. Numerous sites dedicated to representing the body were found (see Figure 10.3B and Figure 10.5C). For reaching, some evidence is suggesting that common reference frames encoded in neurons in the posterior parietal cortex may be used for movement plans (Cohen &

Andersen, 2002), for example. However, 'neurophysiological studies routinely fail to find a significant population of cells whose activity explicitly encodes the output of that transformation in a unique coordinate system. Instead the output may be implicitly embedded in the distributed pattern of activity across the population ... ' (Cisek & Kalaska, 2003; see also Heed et al., 2015). This is perhaps even more prominent in the motor cortex where it is still being debated what it is the cells are specifying about reach—muscle force, movement direction, or a more abstract end goal of muscle action (Lisman, 2015). Somatotopy is rather functional than based on the spatial relationships of the body. The embodied perspective is also appropriate here (see Corbetta, Wiener, Thurman, & McMahon, 2018, for example). At the same time, compared to the octopus, much more of the body seems more explicitly represented. Interestingly, Longo (2015) also considers the implicit–explicit axis for body representations and draws a line roughly between the 'body schema' and the 'body image'. In tasks more related to action and where humans do not consciously represent their body, the body models seem more implicit and also less accurate. These representations may also be dominated by somatosensation and inherit some of the distortions typical of the 'somatosensory homunculi'. Conversely, tasks that relate to conscious perception of our body seem to draw on more explicit representations that are also more accurate/veridical (e.g., image of our hand). This is schematically illustrated in Figure 10.5C.

Finally, works in robotics can also move toward more implicit models—this is in line with the current advent of deep learning and end-to-end architectures. One example from the iCub robot is shown in Figure 10.5B (Nguyen, Hoffmann, Pattacini, & Metta, 2019). From motor babbling experience, the robot learns to associate the head and eye configuration and stereo-image input with arm joint configuration required to reach, together with the body part that will contact the object. The complete mapping is implicitly represented in a deep convolutional neural network. Only the feed-forward component is represented, not the motor execution, which is performed using traditional methods.

### 10.3.4 Serial versus parallel processing

Another axis that separates robot body models from their biological counterparts is serial/sequential versus parallel processing (y-axis in Figure 10.5). This has to do with the design, but also with the computing substrate. Robot control architectures tend to have a serial 'sense–think–act' design; in addition, their control systems run on substrates derived from the classical von Neumann architecture. Biological brains are known for their massively parallel processing. This is nicely illustrated in the example of visually guided behaviour in Figure 10.6 (Cisek & Kalaska, 2003). Interestingly, Pitron et al. (2018) suggest some serial aspect of body representations whereby the body schema would be feeding the body image.

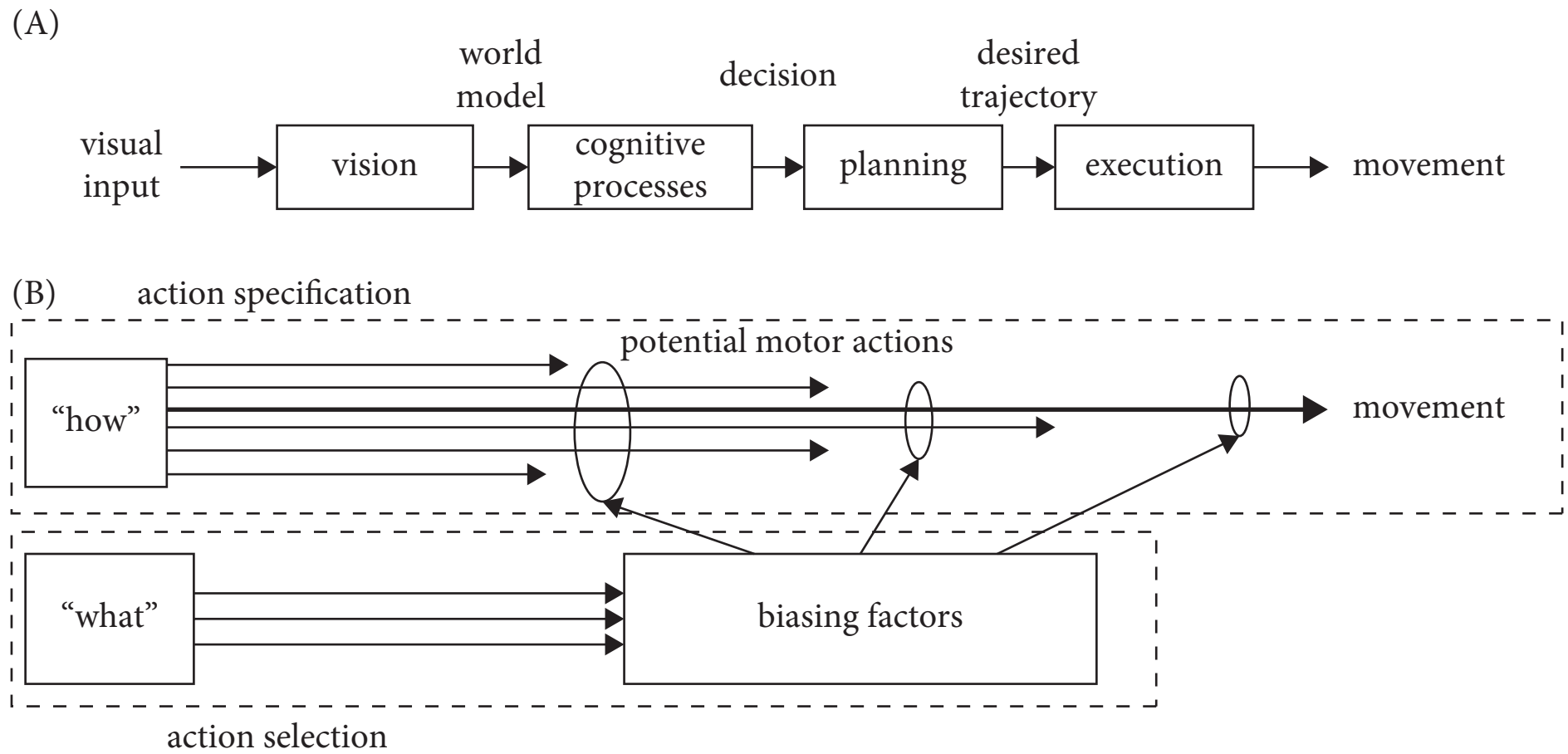

**Figure 10.6** Sequential versus parallel processing for visually guided behaviour. (A) The traditional 'sequential processing' model of visually guided behaviour. In this model, visual input is used to construct a model of the world which is used to make decisions. After decisions are made, a desired trajectory is generated and executed. (B) Schematic representation of the 'specification and selection' architecture for visually guided behaviour. Under this view, visual information has two different roles: specifying the parameters of potential motor actions; and defining criteria which bias competition among those potential actions until a single action is selected. These biasing factors include attention, behavioural relevance, prior reinforcement, required effort, behavioural context, learnt associations, motivations, long-term behavioural objectives, desired outcomes, and any other factor which influences action selection. The processes of specification and selection occur in parallel and continue even during overt movement. A striking feature of this architecture is the absence of a central model of the visual world.

A second aspect is the presence of recurrent connections. In robot control systems, the flow of information is sequential and also unidirectional. Conversely, in neural systems, recurrent connections are ubiquitous. In primate brains, due to their complexity, hierarchies are formed and information flows back and forth, combining the bottom-up and top-down influences at different stages of the hierarchy. This is valid universally, with the circuitry responsible for representing the body not being an exception. The recurrent nature is also more pronounced in primate nervous systems than in cephalopods, say—mainly due to the overall difference in the number of neurons and layers of processing. This difference has also implications on the nature of the online, or short-term, body representations. In traditional robot frameworks, state estimation will be at the top of the sensory processing part, perhaps combined with priors from the model. However, in the brain, state estimation is a highly dynamic, continuous process combining multisensory integration, top-down priors, etc.—through recurrent loops.

Finally, the robot case study (see Figure 10.5B) features a purely feed-forward, hence sequential, neural network.

### 10.3.5 Modular versus holistic and centralized versus distributed representations

A final graphical attempt to contrast body models in robots and animals is depicted in Figure 10.7. At first glance, it would seem intuitive that robot models are centralized, while 'body in the brain' is highly distributed. There is typically only one body model for a robot. In contrast, in the brain, there are numerous distributed, incomplete body representations. In this sense, there is some overlap between centralized versus distributed and explicit versus implicit distinction. However, centralized in this case does not imply monolithic. In fact, robot body models and associated control modules are highly modular, as shown in Figure 10.7A. There are distinct modules like forward/inverse kinematics and dynamics that may draw from the same robot model and be recruited for different purposes like state estimation, movement planning, etc. There would be typically only one module of every kind (imagine a software library) providing this functionality upon request. The representations/modules will thus be universal (as opposed to task-specific) and not overlapping. In nervous systems, on the other hand, there would be often complete sensorimotor loops specialized in a particular task (and hence task-specific rather than universal). In that sense, every such representation would be, in some sense, holistic. Their functionality may also be partially overlapping or redundant. Population coding is a good example; it is striking how unspecific the receptive fields of individual neurons often are (e.g., Georgopoulos, Kalaska, Caminiti, & Massey, 1982). Also, many tasks may require some form of representation of the hand in space, say, but it cannot be excluded that this will be implicitly 'implemented' multiple times. This distributed, specialized, and holistic nature is highest for the nervous system of the octopus (see Figure 10.7C); regarding body models in primates (see Figure 10.7B), the body schema as evolutionarily older and more 'low-level' is on these axes expected to be closer to the octopus, whereas the body image somewhat closer to the engineered body models.

## 10.4 Robots as embodied models of body representations

How can the biological disciplines like cognitive psychology and neuroscience profit from the viewpoint of robot models? The engineering perspective provides a mature analytical machinery that deals with the relevant problems and this can be certainly exploited. For example, consider again the 'multimodal nature of body models' discussed in Section 10.3.2. Next to the Kalman filter metaphor employed by Grush (2004) to describe how estimating the current body state might work, mathematical methods can be also employed to answer fundamental questions like under what exact

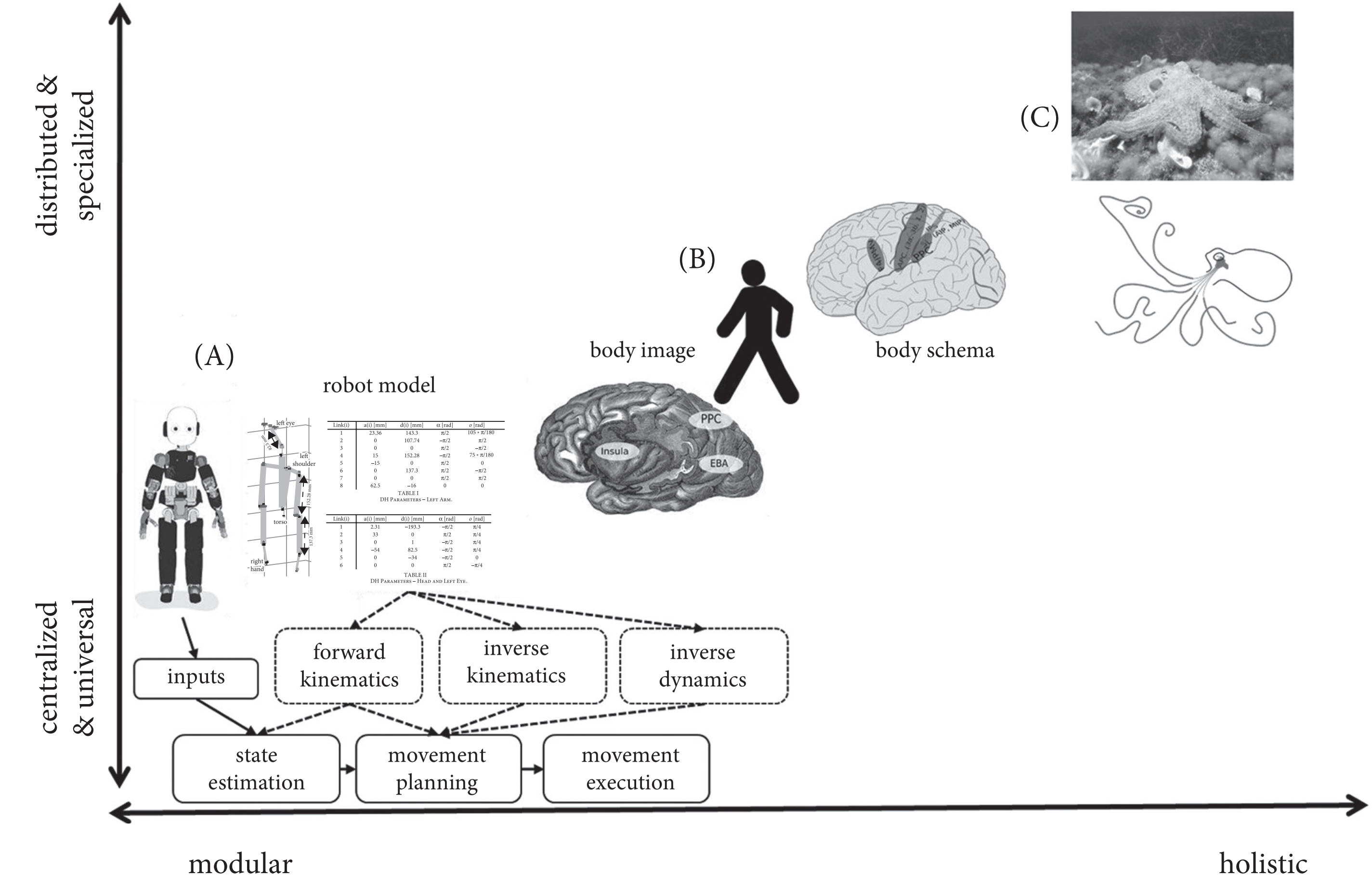

**Figure 10.7** Body representation characteristics III: modular versus holistic and centralized versus distributed. (A) iCub humanoid robot, its kinematic model, and some control modules. (B) Human and schematic illustration of brain areas important for body representations: body schema (see Figure 10.3 for details) and body image (posterior parietal cortex, PPC; extrastriate body area, EBA; insula) (after Berlucchi & Aglioti, 2010). (C) The octopus and its nervous system.

*Image credits*: (A) iCub cartoon: Reproduced courtesy of Laura Taverna, Italian Institute of Technology. (B) Walking human: Public domain (https://commons.wikimedia.org/wiki/File:BSicon_WALK.svg). Body schema brain adapted from Hugh Guiney/ Attribution- ShareAlike 3.0 Unported (CC BY- SA 3.0). Body image brain: Public domain (https://commons.wikimedia.org/wiki/File:Gray731.png). (C) Photo: Common octopus- albert kok/CC BY-SA (https://creativecommons.org/licenses/by-sa/3.0). Octopus nervous system: Jean-Pierre Bellier/CC BY-SA (https://creativecommons.org/licenses/by-sa/4.0).

conditions an agent (or 'the brain') can discover the (amodal) notion of space in which it is embedded and infer its dimensionality from sensorimotor flow only. According to Piaget (1954), for the infant, initially, 'no constant relation exists between visual and buccal space or between tactile and visual space. True, auditory and visual space are already coordinated, as are buccal and tactile space, but no total and abstract space encompasses all the others'. Later, these spaces are connected through prehension (reaching and grasping), and the 'near' and 'far' spaces become differentiated. Eventually, through 'reversible operations'—e.g., whether the object moves in front of me or I move the head, the image on the retina will be the same—the child may overcome the space of individual modalities and 'objectify' the world, space, and its body; the body, say, will appear as an object with certain dimensions, independent of its perception by individual senses. Mathematical and algebraic tools and robotics can formalize Piaget's ideas and provide existence proofs for under what exact conditions an agent may develop spatial knowledge and give precise content to these concepts. Pioneered by Henri Poincaré, compensability (~ reversible operations) was exploited by Philipona, O'Regan, and Nadal (2003) who showed how an agent can infer the dimensionality of space from proprioception and exteroception, and this was extended by Terekhov and O'Regan (2016) to use coincidence detection in neural networks as the basis of a way of discovering the notion of space. Laflaquière, O'Regan, Argentieri, Gas, and Terekhov (2015) explicitly considered the agent's 'point of view' in the sensorimotor flows. An alternative approach makes use of self-contact; the 'body in space' can emerge from the structure of the proprioceptive–tactile space in self-touch configurations (Roschin, Frolov, Burnod, & Maier, 2011; Marcel, Argentieri, & Gas, 2016). Such models do not prove that these solutions are used by the brain, but they provide hypotheses and one can then look for them in the neural code.

Second, next to models at high level of abstraction like the simulated sensorimotor agents, robotics can provide a much higher degree of realism when it comes to mimicking biological bodies. This may be necessary to make progress beyond the existing models addressing coordinate transformations or multisensory integration (Xing & Andersen, 2000; Pouget, Deneve, & Duhamel, 2002) that typically concern very simplified scenarios with one- or two-dimensional geometry, one or two joint angles for the proprioceptive modality, etc. More than 15 years of research in the laboratory of Yasuo Kuniyoshi stands out in this respect; a highly realistic musculoskeletal fetal simulator (21 rigid body parts connected by 20 joints with 36 DoFs, 390 muscles with proprioceptive receptors, and 3000 tactile mechanoreceptor models) has been developed and coupled to a spinal circuit model (neural oscillators, α and γ motor neurons, and sensory interneurons) and a cortical model (2.6 million spiking neurons and 5.3 billion synaptic connections) (Yamada et al., 2016). Figure 10.8A shows the human-like distribution of tactile receptors on the fetal body. Mori and Kuniyoshi (2010) studied the effect of this distribution on the emergence of sensorimotor behaviours; with a natural (non-homogeneous) distribution, the fetus developed 'normal' kicking and jerking movements (i.e., similar to those observed in a human fetus), whereas with a homogeneous allocation, it did not develop any of these behaviours—just one illustration

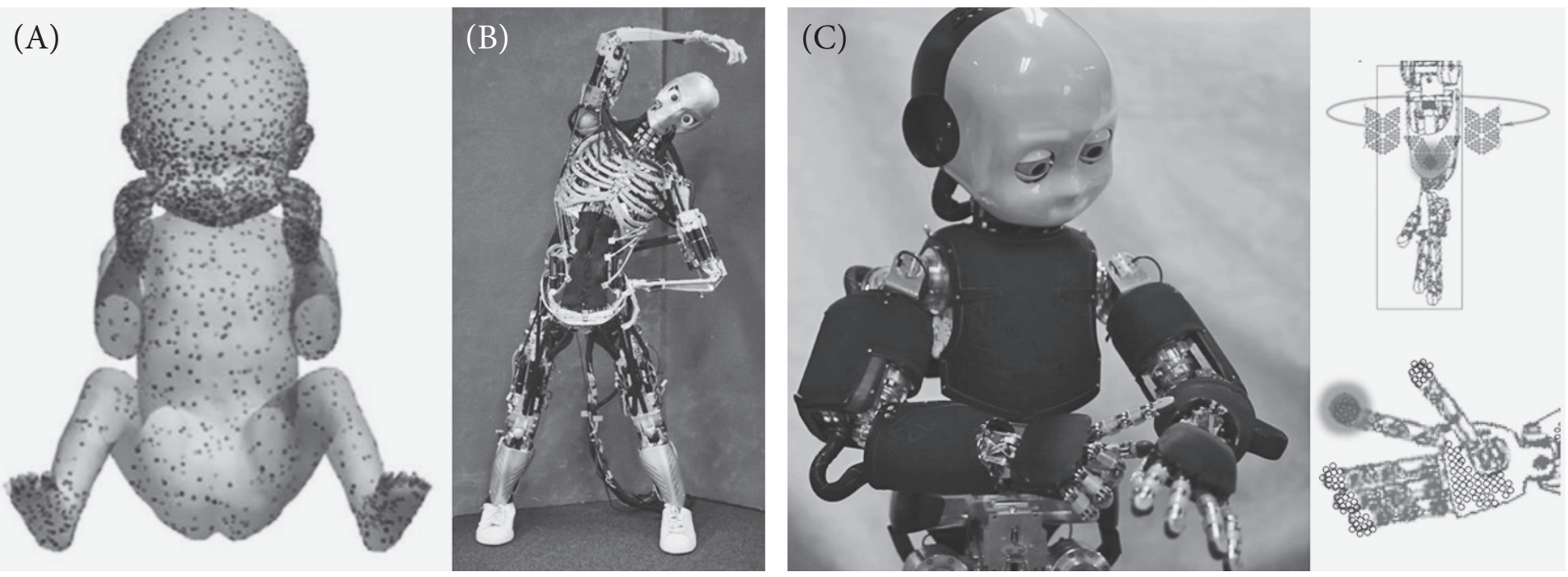

**Figure 10.8** Robots as embodied computational models of body representations. (A) Fetus simulator (Yamada et al., 2016). (B) Musculoskeletal robot Kenshiro (Asano, Okada, & Inaba, 2019). (C) iCub humanoid robot.

*Image credits*: (A) Reproduced from Yamada, Y. et al. An Embodied Brain Model of the Human Foetus, *Sci Rep*, 6 (27893), Figure 1d, doi:10.1038/ srep27893 (2016) under the Creative Commons Attribution 4.0 International License CC BY 4.0 (http://creativecommons.org/licenses/by/4.0/) (B) Reproduced with permission from (Yuki Asano, Kei Okada & Masayuki Inaba, 2019) and courtesy of Yuki Asano and JSK robotics laboratory in the University of Tokyo.

of the importance of the need for embodying the models related to body representations. Simulating physics can be computationally heavy and there is always a risk that the simulator does not get certain properties right. Therefore, physical robots are an indispensable tool. Figure 10.8B shows one of a series musculoskeletal humanoids—Kenshiro (160 'muscles': 50 in the legs, 76 in the trunk, 12 in the shoulder, and 22 in the neck) (Asano, Okada, & Inaba, 2019). Such platforms provide the right challenge to model the impact of the details of the motor system on body representations and reaching behaviours. Richter et al. (2016) have combined a musculoskeletal robotics toolkit (Myorobotics) with a scalable neuromorphic computing platform (SpiNNaker) and demonstrated control of a musculoskeletal joint with a simulated cerebellum. Finally, the iCub baby humanoid robot (see Figure 10.8C) is my platform of choice for models of body representations. It lacks some of the biological details of the other platforms—its whole body skin has a uniform density of tactile receptors and it is driven by standard electric motors rather than by artificial muscles—but it is a very versatile platform with all the key sensory and motor capacities. For example, we showed how it can be used to learn its 'somatosensory homunculus' (Hoffmann, Straka, Farkaš, Vavrečka, & Metta, 2018) or self-calibrate using self-touch (Roncone et al., 2014) or self-touch and self-observation (Stepanova et al., 2019).

## 10.4 Which characteristics of biological body representations should robots take on board?

What properties that biological body representations manifest should robot models copy? One feature that is clearly desirable is adaptivity. Robot models need to be developed in the first place. Then, additional calibration procedures may be required for every robot exemplar and also after deployment on the factory shop floor. During operation, the robot is subject to wear and tear or other conditions might change, calling for additional calibration. All of these processes are costly and often require the intervention of professionals with specialized equipment and possibly suspending production. The 'body in the brain', on the other hand, seems to develop largely from scratch and displays plasticity on all of these timescales—adapting to growth or failures, as well as performing rapid recalibration when working with a tool, for example. Automatic robot self-calibration is thus desired and solutions for this are being developed. For the robot to self-calibrate, it needs redundant sources of information about its body. There is a growing number of powerful, yet economic, sensors for robots available (cameras, RGB-D cameras, inertial measurement units, force/torque sensors, tactile sensors) and they can be exploited for calibration. Multimodality—another property of biological body representations—thus enables plasticity (see Figure 10.4). Such an extension of robot models should thus be unproblematic and is already happening.

I have sketched also other axes—robot body models are typically explicit, veridical (see Figure 10.5), universal, centralized, and modular (see Figure 10.7). All of these are—from an engineering perspective—very convenient properties. For example, being explicit and universal often implies that the models are capable of extrapolation; if transformations in three-dimensional space are represented using appropriate mathematical tools, they will always work—even in previously unseen circumstances. Implicit models would typically be expected to interpolate only, i.e., provide meaningful estimations within the range of existing examples only. Being veridical, or objective, implies that robot body models can be easily validated from the outside. The universal, centralized, and modular nature is ideal from a maintenance perspective. The kinematic model is only in one place and any updates will be automatically propagated to all other modules using it. One important additional convenient property that is a consequence of the features listed above is interpretability; it is possible to understand the model which is key for maintenance, debugging, etc. In this sense, there seem to be good reasons for preserving these properties.

Yet, these very characteristics are responsible for some inherent limitations. In particular, robot body models and associated control architectures lack robustness. The centralized and universal feature makes every module critical and that creates bottlenecks. Redundancy is against software development principles, but it importantly contributes to the resilience of animals. When faced with injuries, impairments, or lesions, they can find alternatives to performing a task. Implicit models are gaining popularity with the advent of deep learning (Nguyen et al., 2019) (for a survey in robotics, see, for example, Sünderhauf et al., 2018). This can be seen as a step toward brain-like models. Making robot control more embodied—exploiting the body morphology or local feedback loops—would be another step in this direction. However, there are trade-offs associated with this; mainly, the interpretability of such implicit, or black-box, models is reduced, which is a downside when they are part of applications where, for example, safety is at stake. Another difference is the sequential versus parallel processing (see Figure 10.6). Having multiple potential movement plans always ready will also improve the robustness of robot behaviour when faced with unexpected situations.

In summary, the strategy employed in robot modelling and control—a single, universal, veridical body model associated with corresponding control schemes—makes robots rather brittle when faced with failures or unexpected changes. Instead, the solution evolved by animals—multiple, distributed, partially overlapping, task-specific, and parallel architectures—makes them particularly robust and resilient.

## 10.5 Conclusion

I have outlined a number of characteristics of robot body models, as well as body representations in nervous systems. Most often, the nature of these models is very different, often ending up on opposite ends of different schematic axes. There seems to be a general trend—on many of the axes sketched in Section 10.3, the sequence is from robot body models, over body image and body schema, to the body representation in the octopus. Paraphrasing Brooks (1991), the octopus is most faithful to the strategy that *the body is its own best model*. Despite the efforts of Brooks and others, robots still heavily rely on models of their bodies—fixed, amodal, explicit, veridical, serial, centralized, and universal. Interestingly, the body image is the second closest to the robot side, which begs the question—asked by Yochai Ataria—to what extent the robots may be like Ian Waterman, the famous 'man who lost his body' (proprioception and touch) (Cole & Paillard, 1995). Indeed, just like deafferented subjects who 'lost their body schema', robots rarely recruit an implicit sensorimotor representation of their body or the body directly (without modelling it) (see also Hoffmann & Müller, 2017). Also, as their body models are explicit and veridical, they resemble the 'pictorial', body image-like representation of their bodies. Vision is also the main sensory modality and most self-calibration approaches rely on it. One difference remains—robots have proprioception, joint angle readings from encoders, and these, together with vision, are employed to bring about reaching movements. However, they rarely have touch (and may be clumsy and slow, just like Ian Waterman) and without their body model cannot do pretty much anything.

We have seen that the way animals and machines represent their bodies is quite different. Can robots contribute to our understanding of the 'body in the brain'? As we have seen in Section 10.4, this can be the case in two ways: (1) employing the mathematical machinery can provide proofs of what is possible—extracting body as an object in three dimensions from multimodal sensory information, for example; and (2) using robots as embodied computational models of body representations. In all cases, one should reflect upon using robots, such that the 'design decisions' typical for robot models are not blindly applied to the biological world. In terms of performance (Section 10.5), interestingly, the rather 'messy' way adopted by biology—which is also why our understanding of the mechanisms is still limited—is surprisingly good. Neither the models nor the behavioural performance are optimal, but they are good enough and highly robust. On the other hand, robot body models and control are very neat, transparent, universal, and overall highly parsimonious and optimized. Yet, the performance is fragile. One factor to consider here is that the criterion for quality of the models in animals is the success in the action/task for which the particular body representations were recruited. Robot body models ultimately also serve actions; however, the criterion for model quality is typically rather that it is a veridical representation of something—the robot link length, for example. The deep learning type of models may provide one of the ways of bridging machine learning and neuroscience (e.g., Richards et al., 2019).

## Acknowledgements

This work was supported by the Czech Science Foundation (GA CR), project EXPRO (nr. 20-24186X). The chapter draws partly on an earlier manuscript in preparation and discussions with Julius Verrel in 2012. I would like to thank the editors Yochai Ataria, Shogo Tanaka, and Shaun Gallagher for their comments and patience. I am also indebted to Kevin O'Regan, Alban Laflaquiere, and Christian Mangione for comments on a draft of the manuscript, and Gregor Schöner for discussions on the development of reaching. Yochai Ataria drew my attention to the parallel between robots and Ian Waterman. Kevin O'Regan pointed out the 'body in the brain' versus 'brain in the body' implications.